\documentclass{cimento}

\usepackage{graphicx}  

\title{Negative heat capacity for hot nuclei: confirmation with
formulation from the microcanonical ensemble}
\shorttitle{Negative heat capacity for hot nuclei: confirmation with {\ldots} }
\author{B.~Borderie\from{ins:1}\thanks{E-mail:
bernard.borderie@ijclab.in2p3.fr}\ETC,
S.~Piantelli\from{ins:2},
J.D.~Frankland\from{ins:3}
        \atque
N.~Le Neindre\from{ins:4}}
\instlist{\inst{ins:1} Universit\'e Paris-Saclay, CNRS/IN2P3, IJCLab,
91405 Orsay, France
  \inst{ins:2} INFN, Sezione di Firenze, Sesto Fiorentino, Italy
  
  \inst{ins:3} GANIL, UPR 3266, CEA-DRF/CNRS-IN2P3), 14076 Caen, France
  
  \inst{ins:4} Normandie Univ., ENSICAEN, UNICAEN, CNRS/IN2P3, LPC,
  14050 Caen, France
  }

\begin{document}

\maketitle

\begin{abstract}
By using freeze-out properties of multifragmenting hot nuclei produced
in quasifusion central $^{129}$Xe+$^{nat}$Sn collisions at different 
beam energies 
(32, 39, 45 and 50 AMeV) which were estimated by means of a
simulation based on experimental data collected by the $4\pi$    
INDRA multidetector, heat capacity in the thermal excitation energy
range 4 - 12.5 AMeV
was calculated from total kinetic energies and multiplicities at
freeze-out. The microcanonical formulation was employed.
Negative heat capacity which indicates a first order phase transition
for finite systems is observed and confirms previous results using
a different method. 
\end{abstract}

\section{Introduction}
An important challenge of heavy-ion collisions at intermediate
energies was the highlighting and characterization of the 
liquid-gas phase transition in hot nuclei. 
At present huge progress has been made even if some points can be deeper
investigated~\cite{WCI06,Bor08,Bor19}. This was notably the case for
the observation of negative microcanonical heat
capacity related to the consequences of local convexity of the entropy
for finite systems~\cite{Cho04,Bor19}.

About twenty years ago MULTICS and INDRA collaborations highlighted
this signal of negative heat capacity~\cite{MDA00,NicPHD,Bor02}.
The method to derive heat capacity was proposed in~\cite{Cho99} and
applied to both experimental and microcanonical lattice gas 
model~\cite{Cho00} data showing for the model that negative heat
capacity appeared as a robust signal. The method is based on the fact
that for a given total thermal energy, the average partial energy
stored in a part of the system is a good microcanonical thermometer,
while the associated fluctuations can be used to construct heat
capacity. In this approach a single temperature is used to describe
the system at freeze-out: the same temperature 
is associated with both internal excitation and thermal motion 
of emitted fragments, which is not physically obvious if one 
remembers that the level density is expected to vanish at high 
excitation energies~\cite{Tub79,Dean85,Koo87}.
On the other hand it was also shown a few years after, from a
detailed simulation, the necessity to impose a limitation for the
temperature of fragments to be able to reproduce experimental data 
and
consequently the necessity to use two temperatures: a microcanonical
temperature corresponding to the thermal motion and a second temperature
corresponding to the internal excitation of fragments. Exact
microcanonical formulae with the two temperatures were proposed 
in~\cite{Radut02,Radut00}; they are used in this work. Both methods
need information from data reconstructed at freeze-out.

In the paper we shall first recall previous results with the first method
called method 1 in what follows. Then a section will be devoted to
the detailed simulation for reconstructing freeze-out properties. Results
applying exact microcanonical formulae (method 2) are presented 
in section 4. Finally before concluding results with both methods are discussed.

\section{Microcanonical heat capacity with method 1 (partial
energy fluctuations)}
The method proposed in~\cite{Cho99} was applied to quasi-fused (QF) 
systems for central $^{129}$Xe+$^{nat}$Sn collisions at different 
bombarding energies: 32, 39, 45 and 50 AMeV. Experimental data were 
collected with the 4$\pi$ multidetector INDRA described in detail 
in refs.~\cite{I3-Pou95,I5-Pou96}.
Accurate particle and fragment identifications were achieved
and the energy of the detected products was measured
with an accuracy of 4\%. Further details can be found 
in refs.~\cite{I14-Tab99,I34-Par02}. 

Without entering into all the details of method 1, we can just recall
the main points. From experiments the most simple decomposition of the thermal
excitation energy 
is in a kinetic part, $E_{k}$, and a potential part, $E_{pot}$ 
(Coulomb energy + total mass excess).
These quantities have to be determined at freeze-out  and
consequently it is necessary to trace back this
configuration on an event by event basis.
The true configuration needs the knowledge of the freeze-out
volume and of all the particles evaporated from primary hot
fragments including the (undetected) neutrons. Consequently
some working hypotheses are used, constrained by
specific experimental results (see for example~\cite{MDA02}).
Then, the experimental correlation between the kinetic energy per nucleon
$E_{k}$/$A$ and
the thermal excitation energy per nucleon $E^{*}$/$A$ of the considered
system can be obtained event by event 
as well as the variance of the kinetic energy
$\sigma_{k}^{2}$. Note that $E_{k}$ is calculated by subtracting 
the potential part $E_{pot}$ from the thermal excitation energy $E^{*}$ 
and consequently
kinetic energy fluctuations at freeze-out reflect the configurational 
energy fluctuations.
An estimator of the
microcanonical temperature of the system can be obtained from the
kinetic equation of state:
\begin{equation}
\noindent < E_{k} > = \langle \sum_{i=1}^{M} a_i \rangle T^2 +
\langle \frac{3}{2} (M-1) \rangle T 
\label{eq:Keos}
\end{equation}
The brackets $\langle\rangle$
indicate the average on events with the same $E^{*}$,
$a_i$ is the level density parameter and M the multiplicity at
freeze-out. \textit{In this expression the same
temperature is associated with both internal excitation and thermal
motion of fragments.}
Then an  estimate of the total microcanonical heat capacity 
is extracted using three equations.
\begin{equation}
\noindent C_k = \frac{\delta <E_k / A >}{\delta T},
\label{eq:CnegM1}
\end{equation}
is obtained by taking the derivative of $<E_k / A >$ with 
respect to $T$.
\begin{figure}
\begin{center}
\includegraphics[width=0.24\textwidth]{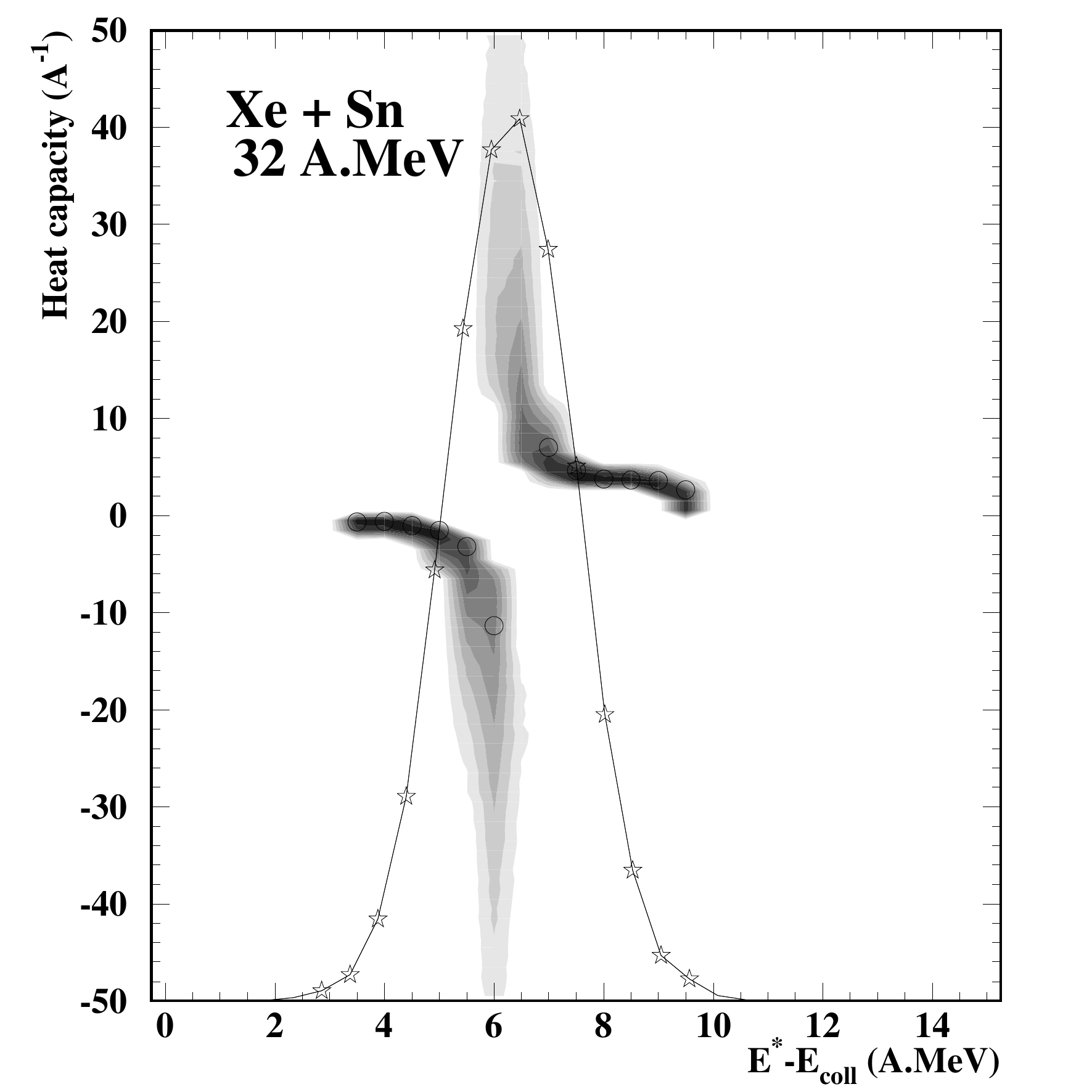}
\includegraphics[width=0.24\textwidth]{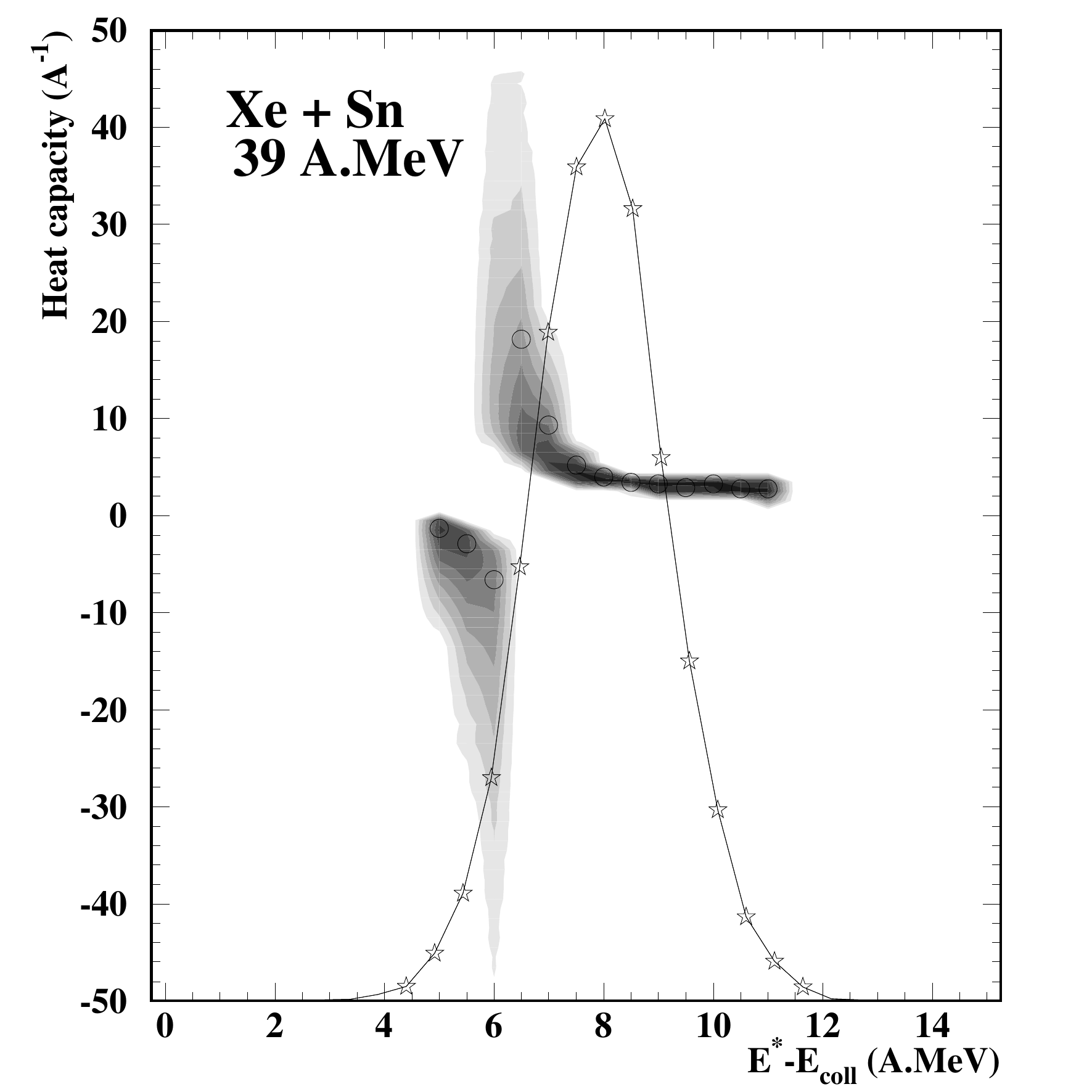}
\includegraphics[width=0.24\textwidth]{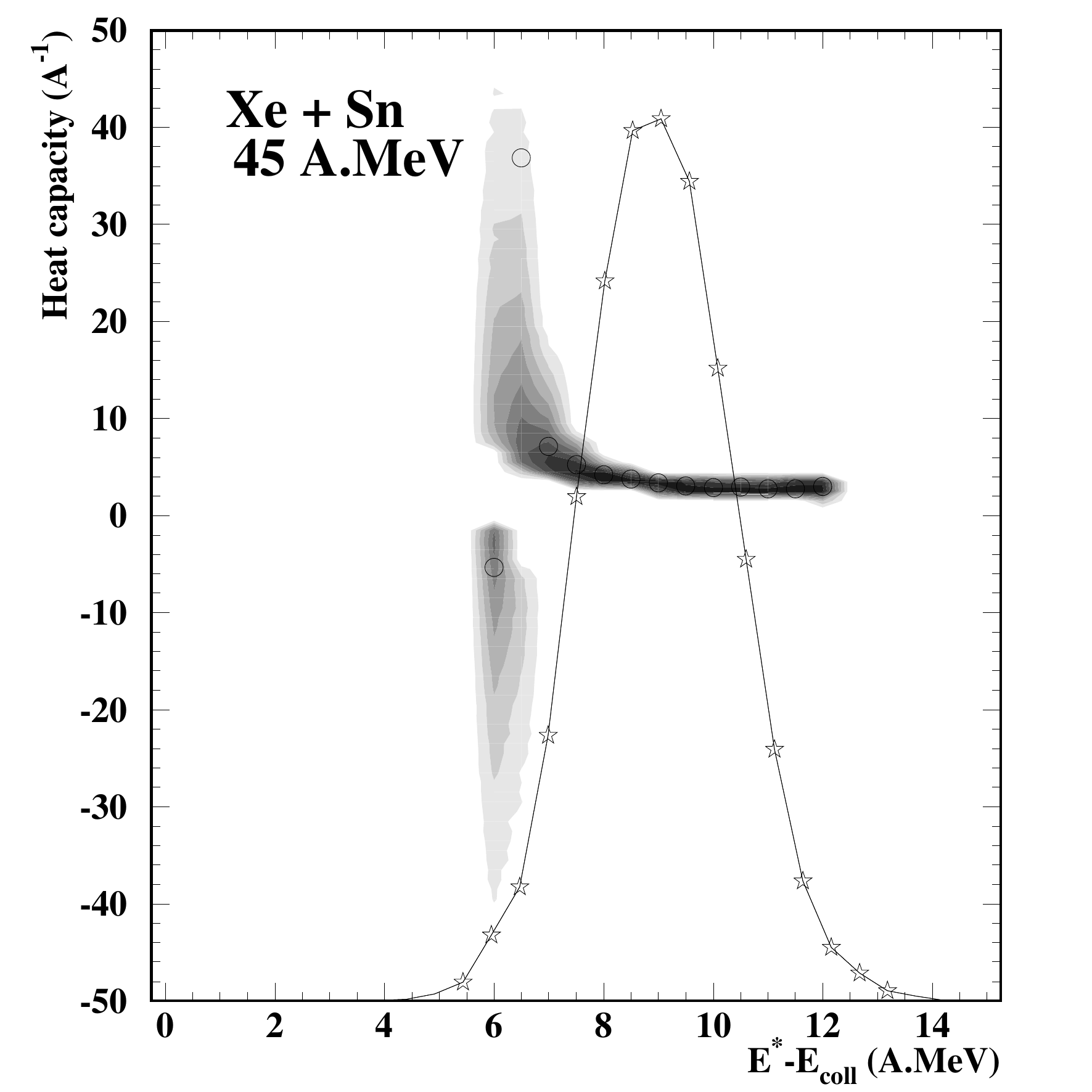}
\includegraphics[width=0.24\textwidth]{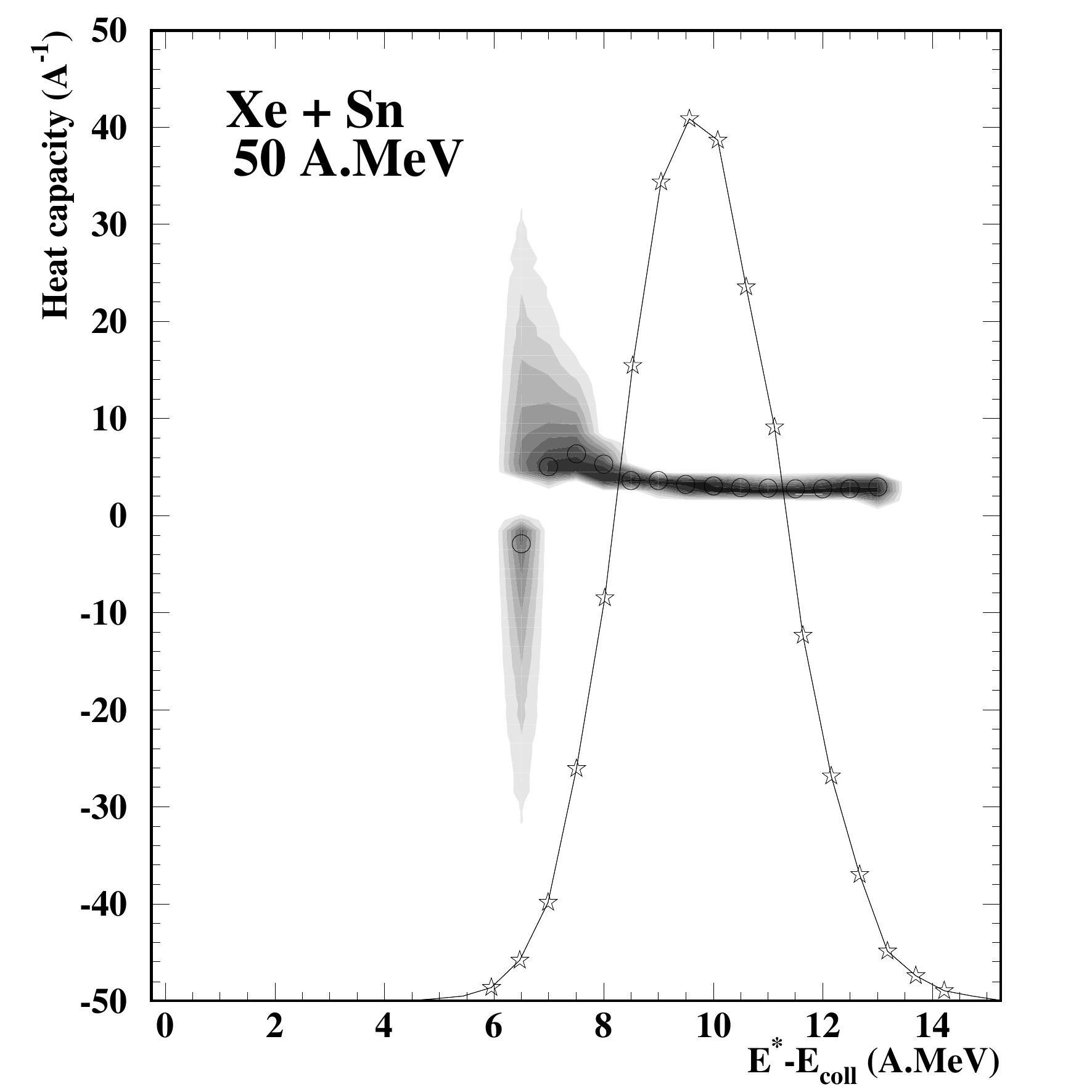}
\end{center}
\caption{
Microcanonical heat capacity per nucleon as a function of the thermal
excitation energy (total excitation energy corrected on average for
the radial collective energy).}
\label{fig:meth1}
\end{figure}

Using a Gaussian approximation for the kinetic energy distribution
its variance can be calculated as\\
\begin{equation}
A\sigma_{k}^{2} \simeq T^2\frac{C_kC_{pot}}{C_k+C_{pot}}.
\label{eq:CnegM2}
\end{equation}
Eq.~(\ref{eq:CnegM2}) can be inverted to extract, from the observed
fluctuations, an estimate of the microcanonical heat capacity:
\begin{equation}
\noindent (\frac{C}{A})_{micro} \simeq C_k+C_{pot} \simeq \frac{C_k^2}
 {C_k - \frac{A \sigma_k^2}{T^2}}.
\label{eq:CnegM3}
\end{equation}
From Eq.~(\ref{eq:CnegM3}) we see that the specific microcanonical
heat capacity $(C/A)_{micro}$ becomes negative if the normalized
kinetic energy fluctuations $A \sigma_k^{2}/T^2$ overcome $C_k$.
Figure ~\ref{fig:meth1} shows the 
results obtained~\cite{NicPHD,Bor02}. The heat capacity
is plotted as grey zones (error bars). At 32 and 39 AMeV a negative
branch is observed.

\section{A detailed simulation for reconstructing freeze-out
properties - the necessity to use two temperatures}
Starting from the same raw data
the reconstruction of freeze-out
properties from simulations~\cite{I66-Pia08,I58-Pia05} was the following. 
Data with a very high degree of charge completeness were selected, 
(measured fraction of the available charge $\geq$93\% 
of the total charge of the system), which is crucial 
for a good estimate of Coulomb energy. QF sources were 
reconstructed, event by event,
by summing the contributions of fragments (Z$\geq$5) at all angles and
doubling that of light charged particles (Z$<$5) emitted between
60 and 120$^{\circ}$ in the reaction centre of mass, in order to exclude
the major part of pre-equilibrium
emission~\cite{I69-Bon08,I29-Fra01}; with such a prescription only
light charged particles
with isotropic angular distributions and angle-independent average  
kinetic energies are considered. 
\begin{figure}
\begin{center}
\includegraphics[width=0.80\textwidth]{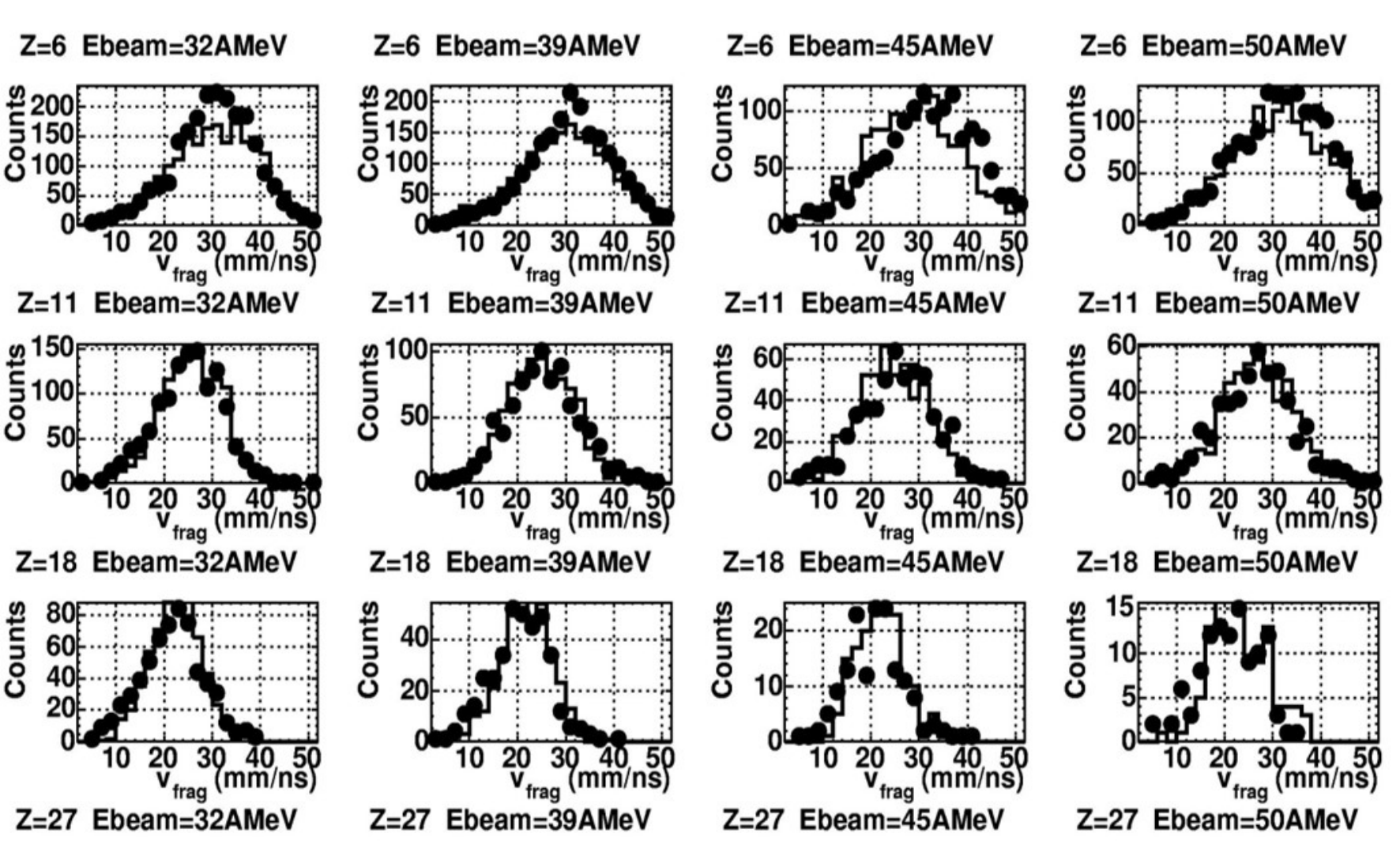}
\end{center}
\caption{
Comparison in the reaction centre of mass between the experimental 
velocity spectra of fragments of a given charge (full points)
and the simulated ones (histograms). Each
row refers to a different fragment charge and each column to a different
beam energy. From~\cite{I66-Pia08}.}
\label{fig:distfrag}
\end{figure}
In simulations,
dressed excited fragments and particles at
freeze-out are described by spheres at normal density.
Then the excited fragments subsequently deexcite while flying apart.
All the available asymptotic experimental information (charged
particle spectra, average and standard deviation of fragment velocity
spectra and calorimetry) is used to constrain the four free parameters
of simulations to recover the data at each incident energy: the percentage
of measured particles which were
evaporated from primary fragments, the collective radial energy, a minimum
distance between the surfaces of products at freeze-out and a limiting
temperature for excited fragments found equal to 9 MeV. All the details of simulations can be found
in refs.~\cite{I66-Pia08,I58-Pia05}.
The limiting temperature,
related to the vanishing of level density for fragments~\cite{Koo87},
was found mandatory to reproduce the observed widths of fragment velocity
spectra. With a single temperature (internal and kinetic temperatures
equal) the sum of Coulomb repulsion, collective energy, 
thermal kinetic energy directed at random
and spreading due to fragment
decays accounts for about 60-70\% of those widths.
By introducing a limiting temperature, 
which corresponds to intrinsic temperatures for fragments
in the range 4-7 MeV (see figure 1 of~\cite{I79-Bor13}), 
the thermal kinetic
energy increases, due to energy conservation, thus producing the missing
percentage for the widths of final velocity distributions.
As shown in fig.~\ref{fig:distfrag}, the agreement between experimental 
and simulated velocity spectra for fragments, for the
different beam energies, is quite remarkable. 

\section{Direct formulae from the microcanonical ensemble - method 2}
Direct formulae have been proposed in ref.~\cite{Radut02}
to calculate heat capacity but never used to extract information from data.
They are derived within the microcanonical ensemble
by considering fragments interacting 
only by Coulomb and excluded volume,
which corresponds to the freeze-out configuration.
Within this ensemble, the statistical weight of a configuration
$c$, defined by the mass, charge and internal excitation energy
of each of the constituting $M_c$ fragments,
can be written (see~\cite{Radut02,I94-Bor20}).
To apply the deduced formulae for microcanonical temperature and second
derivative of the system entropy versus thermal energy or alternatively  heat
capacity, two parameters were fixed in the microcanonical ensemble to 
ensure coherence with the simulations performed to estimate quantities at 
freeze-out. These are the fragment level density in which the limiting
temperature for fragments is fixed at 9 MeV~\cite{Radut00} as 
obtained from simulations 
and the number of kinetic degrees of freedom which was fixed according to the
conservation of energy and linear momentum as in simulations. 
The microcanonical temperature is deduced from its statistical
definition~\cite{Radut02}:
\begin{eqnarray}
T=\left(\frac{\partial S}{\partial
E^*}\right)^{-1}&=&\left(\frac1{\sum_c W_c} \sum_c 
W_c(3/2M_c-5/2)/K\right)^{-1}=\langle(3/2M_c-5/2)/K\rangle^{-1}
\label{eq:T}
\end{eqnarray}
$W_c$ is the statistical weight of a configuration and
the notation $\langle\rangle$ refers to the average over the ensemble
states. The heat capacity of the system, $C$, is related to
the second derivative of the entropy by the equation
$\partial^{2}S /\partial E^{*2}$ = $-1/CT^{2}$.
Thus, one can evaluate the second derivative of the system entropy
versus $E^*$~(Eq.~(\ref{eq:15})) or alternatively the heat 
capacity $C$~(Eq.~(\ref{eq:14})).
\begin{eqnarray}
\frac{\partial^{2} S}{\partial
E^{*2}}&=&\langle \frac{(3/2M_c-5/2)(3/2M_c-7/2)}{K^2}\rangle
-\langle\frac{(3/2M_c-5/2)}{K}\rangle^2
\label{eq:15}
\end{eqnarray}

\begin{eqnarray}
C=\left(1-T^2\langle \frac{(3/2M_c-5/2)(3/2M_c-7/2)}{K^2}\rangle\right)^{-1}
\label{eq:14}
\end{eqnarray}
These two quantities only depend on two parameters, $M_c$, the total
multiplicity and , $K$, the total thermal kinetic energy 
estimated at freeze-out from the detailed simulation.
\begin{figure}
\begin{center}
\includegraphics[width=0.30\textwidth]{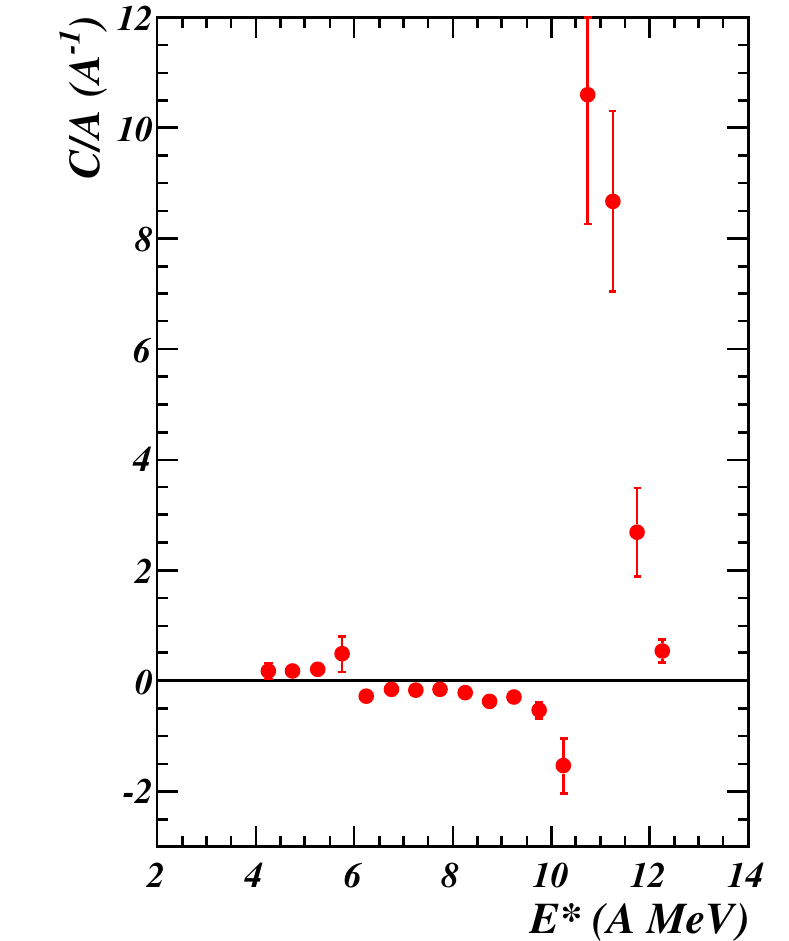}
\includegraphics[width=0.30\textwidth]{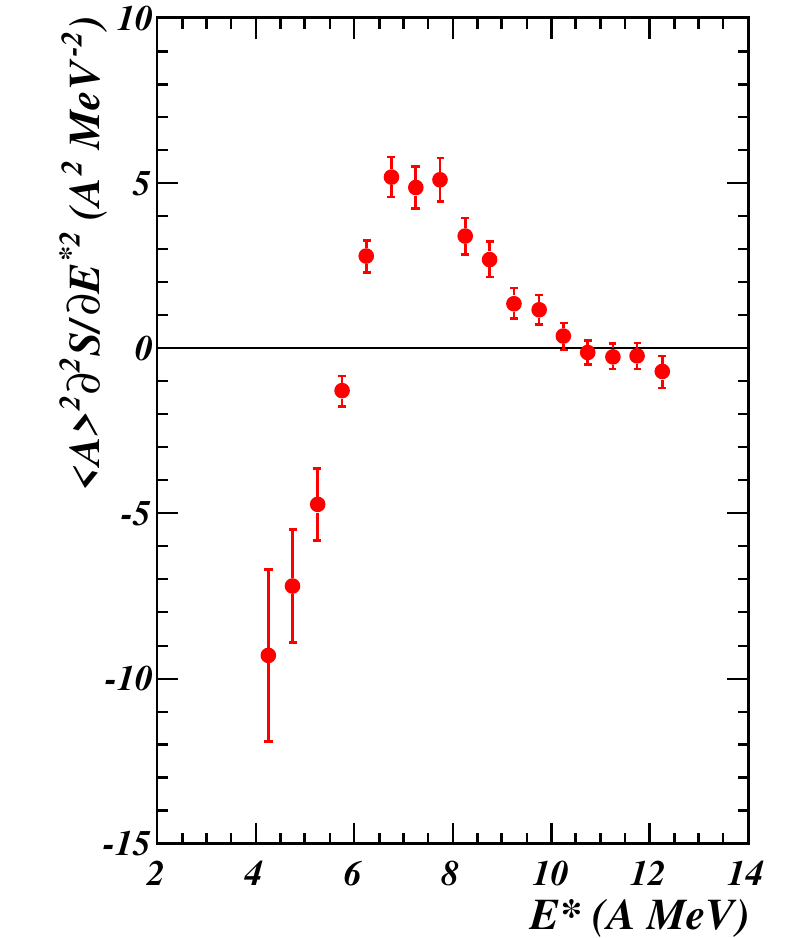}
\end{center}

\caption{ Heat capacity (left) and second derivative of the 
entropy (right) versus thermal excitation energy $E^*$. 
Error bars correspond to systematic plus statistical errors (see text).
From~\cite{I94-Bor20}.}
\label{fig:cneg}
\end{figure}

Values of heat capacity and second derivative of the entropy versus thermal 
excitation energy $E^*$
have been calculated respectively from Eqs.~(\ref{eq:14}) and~(\ref{eq:15}) for QF 
hot nuclei
with $Z$ restricted to the range 80-100 to suppress tails of the 
distributions and by putting together simulation results from the
different incident energies.
The average over the ensemble states have been assimilated to an
average over ``event ensembles'' sorted into $E^*$ bins.
A binning of 0.5 AMeV was chosen to have a sufficient number
of events in each bin in order to reduce statistical errors.
Figure~\ref{fig:cneg} shows the results.
Error bars correspond to systematic plus statistical
errors; systematic errors were evaluated by varying the free parameters
of simulations within their limits defined by a $\chi^2$ procedure~\cite{I66-Pia08}.
The left part of the figure shows the results for the direct calculation 
of $C/A$. Negative heat capacity is observed on a
rather large thermal excitation energy range and
the second diverging region is more visible than the first one.
As the second derivative of the entropy is a very small 
quantity (from $\sim 2.10^{-4}$ to $\sim 3.10^{-6}$), 
we have kept the presentation made in~\cite{Radut02} i.e.
$A^{2} \partial^{2}S /\partial E^{*2}$ for fig.~\ref{fig:cneg} - right
part; A is replaced by the average mass of the QF hot
nuclei, $<A>$, on the considered
$E^*$ bin. This quantity
better defines the $E^*$ domain of negative heat capacity.  
Positive values are measured in the range 6.0 - 10.0 AMeV.
The related microcanonical temperatures calculated with 
Eq.~(\ref{eq:T}) are displayed in fig.~\ref{fig:CnegTmTs};
they are rather constant around
17 - 18 MeV in the $E^*$ range where negative heat capacity 
is observed. With the large multiplicities observed in the present 
study, microcanonical temperatures are close to classical kinetic temperatures
(see figure 3 of~\cite{I79-Bor13}).

\section{ The two methods - comparison of results}
As compared to heat capacity estimates of method 1~\cite{NicPHD,Bor02}, 
there is a significant difference with results of method 2 for the
$E^*$ range of negative values:  6.0$\pm$1.0 - 10.0$\pm$1.0 AMeV for 
method 2 and $<$4.0$\pm$1.0 - 6.0$\pm$1.0 AMeV with method 1.
For QF hot nuclei selection the same event shape sorting was
used. The degree of completeness was different (93\% here to be compared to
80\% before) but it does not affect significantly the thermal
excitation energy per nucleon. 
\begin{figure}
   \begin{minipage}[t]{0.32\textwidth}
      \includegraphics[width=\textwidth]{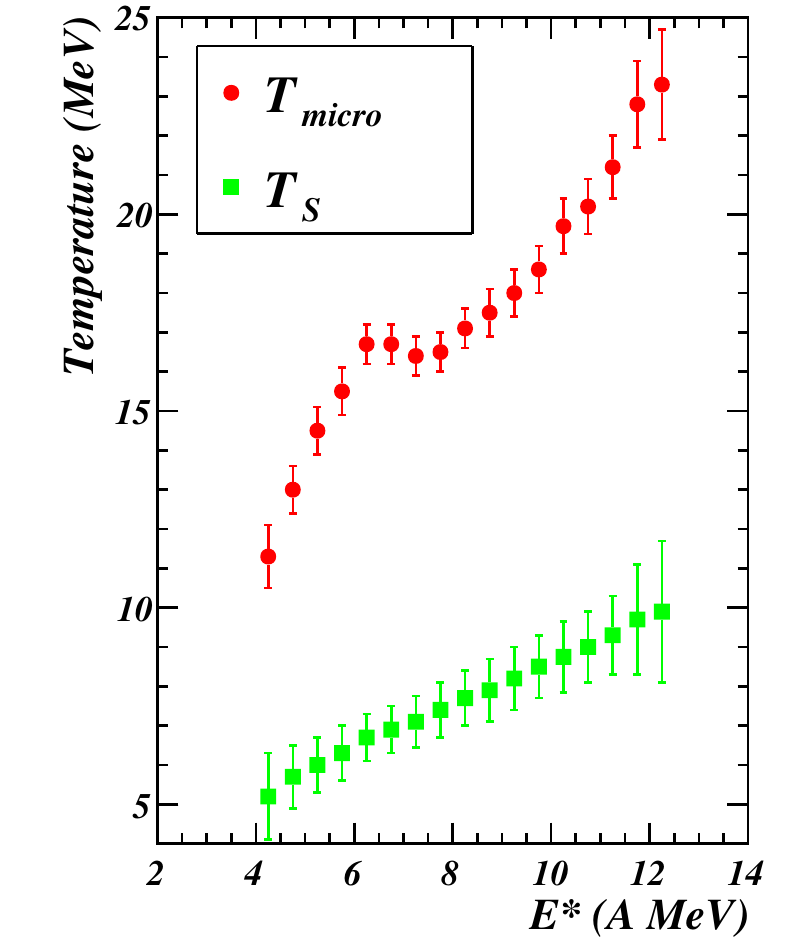}
      \caption{Evolution of temperatures with thermal excitation energy:
      microcanonical temperature , $T_{micro}$, used in method 2 and
      single temperature, $T_S$, for method 1. Error bars correspond to 
      systematic plus statistical errors.}
      \label{fig:CnegTmTs}
      \end{minipage}%
   \hspace*{0.2\textwidth}%
   \begin{minipage}[t]{0.32\textwidth}   
       \includegraphics[width=\textwidth]{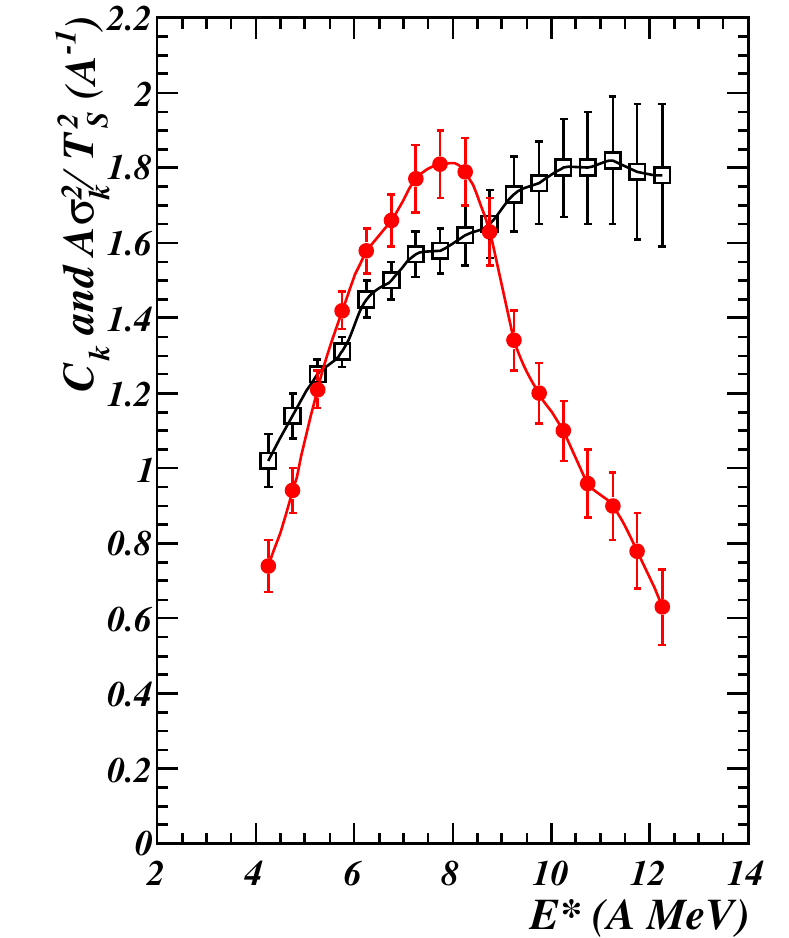}
      \caption{Normalized kinetic energy fluctuations (filled circles) and
       estimated $C_k$ values (open squares) related to
       method 1 (see text). Error bars correspond to systematic plus 
       statistical errors. From~\cite{I94-Bor20}.}
      \label{fig:Cnegapp}    
   \end{minipage}           
\end{figure}
The method to reconstruct
freeze-out properties was also different.
But the main difference seems to be related to the average freeze-out 
volume. In method 1 the
average freeze-out volume used was kept constant at 3 times the volume at
normal density (3 $V_0$) over the whole thermal excitation energy range
whereas with the detailed simulation it varies 
from 3.9 to 5.9 $V_0$ (see fig. 4 of ~\cite{I94-Bor20}).
To check this, method 1 has been applied to freeze-out data 
used with method 2.
From the detailed simulation, for each $E^*$ bin, we have calculated $<E_k>$
(see Eq.~(\ref{eq:Keos})) and derived an apparent single temperature,
$T_S$ (see fig.~\ref{fig:CnegTmTs}), needed to build the normalized 
kinetic energy fluctuations, $A\sigma_{k}^{2}$/$T_{S}^{2}$, 
to be compared to $C_k$ (see Eqs.~(\ref{eq:CnegM1}) and~(\ref{eq:CnegM3})). 
Figure~\ref{fig:Cnegapp} shows that heat capacity becomes negative in
the $E^*$ range 5.5$\pm$1.0 - 9.0$\pm$1.0 AMeV, i.e. when
$A\sigma_{k}^{2}$/$T_{S}^{2}$ overcomes $C_k$.
This clearly confirms that the main difference, as compared to 
estimates with method 1, comes from different average freeze-out volumes.
We also note a small decrease of the $E^*$ domain of negative heat
capacities as compared to
method 2, which possibly comes from approximations made in method 1.

\section{Conclusion}
Heat capacity measurements have been revisited without
approximation and by correcting the hypothesis of a single temperature
associated with both internal excitation and thermal motion of fragments.
For those measurements microcanonical formulae and data
reconstructed at freeze-out with the help of a detailed simulation have 
been used. Negative heat capacity was confirmed for hot nuclei 
in the coexistence region of the liquid-gas phase transition.
For the future, one of the last points to be also deeper investigated
is the bimodality signature for QF hot nuclei. As demonstrated for
$\Delta$-scaling, the effect of the onset and increase of radial 
expansion must be understood~\cite{Bor19,I51-Fra05,I78-Gru13,I69-Bon08}.


\end{document}